\documentclass[runningheads,a4paper]{llncs}
\usepackage{cite}
\usepackage{bbding}
\usepackage{algorithmic}
\usepackage{graphicx}
\usepackage{hyperref}
\usepackage{endnotes}
\usepackage{url}
\usepackage{longtable}
\usepackage{xcolor}
\def\BibTeX{{\rm B\kern-.05em{\sc i\kern-.025em b}\kern-.08em
    T\kern-.1667em\lower.7ex\hbox{E}\kern-.125emX}}

\begin{document}

\title{Connected Vehicles: A Privacy Analysis}

\author{Mark Quinlan \inst{ (}\Envelope\inst{)} \and
Jun Zhao\ \and Andrew Simpson}

\institute{Department of Computer Science, University of Oxford\\
Wolfson Building, Parks Road, Oxford OX1 3QD, United Kingdom\\
\email{firstname.lastname@cs.ox.ac.uk}\\}

\maketitle      


\begin{abstract}
Just as the world of consumer devices was forever changed by the introduction of computer controlled solutions, the introduction of the engine control unit (ECU) gave rise to the automobile's transformation from a transportation product to a technology platform.  A modern car is capable of processing, analysing and transmitting data in ways that could not have been foreseen only a few years ago.  These cars often incorporate telematics systems, which are used to provide navigation and internet connectivity over cellular networks, as well as data-recording devices for insurance and product development purposes.  We examine the telematics system of a production vehicle, and aim to ascertain some of the associated privacy-related threats.  We also consider how this analysis might underpin further research.
\end{abstract}

\section{Introduction}

A modern automobile equipped with systems for navigation and communication, such as a wireless modem, is generally called a \emph{connected car}.  Such cars also have various systems connected to an in-vehicle network (often called the Controller Area Network bus, or CAN-Bus --- a message protocol that allows multiple micro-controllers on a network to communicate without a single host computer) that collect data on their usage.  

Naturally, connected vehicles bring with them worries relating to the privacy of personal data.  A recent example of a connected vehicle privacy issue is the illegal tracking of leased vehicles in France\endnote{\url{http://www.lepoint.fr/high-tech-internet/geolocalisation-la-cnil-rend-}\\\url{une-sanction-exemplaire-31-07-2014-1850400\_47.php}}. In that case, it was found that the company was installing an additional tracking unit onto the CAN-Bus, which would relay not only GPS coordinates, but also vehicle usage statistics~---~without the user's knowledge or consent.  Furthermore, connected cars present a significant opportunity for data misuse.  For example, users might fabricate their location or usage data, or use the built-in applications maliciously~\cite{Atzori}, while, on the manufacturer side, there are opportunities to share or sell data to third parties without appropriate consent.

Many of the systems currently in place do not allow for significant user control over what kinds of data are collected, nor do they have clear privacy policies in place~\cite{weber}. In many cases, there may be users who are not aware of their data being collected and used by third parties~\cite{Larkin}, or even that a privacy policy for their vehicle exists~\cite{surveyonsecurity}. 

We provide a high-level overview of the privacy risks affecting the current connected vehicle landscape. To this end, we provide a high-level assessment of the threat landscape based on an examination of a telematics unit, and extract sample data. We then make inferences to privacy issues surrounding the larger data transmission, handling and storage infrastructure.

\section{Background}\label{motivation}

Within the Internet of Things (IoT) landscape, significant attention has been focused on privacy aspects relating to the use of connected objects. While continuously connected smartphones have been a consistent topic of interest~\cite{privacy1,privacy2}, connected cars research has tended to focus on security (e.g.~\cite{hacktarget}).

Contributions such as~\cite{Hubaux} and~\cite{othmane} provide foundations with respect to security; they also provide a background for privacy concerns, without directly assessing a production telematics system for such threats.  More pertinently: \cite{glancy} provides an overview of the expectations and interests of the users and developers of connected vehicles; \cite{lim} expounds on the use of potentially nefarious use of location-based services as a privacy threat; and~\cite{kaplun} illustrates how data generated by connected vehicles can be used for usage-based insurance purposes.

Our primary concern is privacy: 
we do not concern ourselves with security flaws (other than when such flaws lead to privacy compromises).  A key concern has been an analysis of the data that these devices explicitly capture and return to their manufacturers.  We gave consideration to an analysis of a popular telematics systems produced by a global manufacturer.  A policy review was conducted, which yielded information pertaining to general areas of data collected.

\section{Data acquisition}

\begin{figure*}[t]
\centering

\includegraphics[scale=0.35]{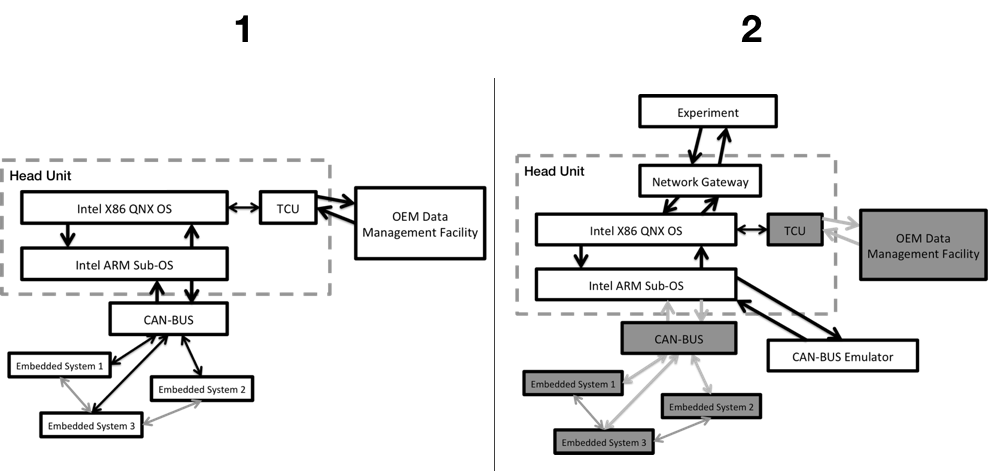}

\caption{The architecture as understood at a high-level is shown on the left-hand side.  The right-hand side highlights the entry-method that was used. The sections of the original system that needed to be bypassed have been greyed out.}
\label{Fig.1}
\end{figure*}

\subsection{Choice of unit}

We considered a connected vehicle telematics unit (by which we mean a head unit and/or a head unit with a TCU (Telematics Communications Unit, which we subsequently refer to as `the sample-unit') featuring built-in internet connectivity that can be used without prior user set-up).  With respect to our chosen manufacturer, any vehicle from 2009 onwards fitted with either a head unit and modem combination or a head unit with a built-in modem unit met the definition.  

Our sample-unit was taken from a 2014 vehicle, of which the technology powering it can be found within production vehicles today, and chosen due to its similarity in terms of functionality with units provided by other manufacturers.  The sample-unit was chosen on the basis of the following.
\begin{itemize}
\item A system built upon QNX\footnote{\url{http://www.qnx.com/news/events/japan-summit/en/presentations/Connectivity\%20in\%20automotive_en.pdf}} was a desirable feature, due to QNX being a commercial Unix-like real-time operating system developed by Blackberry that has been used in over 60 million cars (and other products, such as tablets and mobile phones)\footnote{\url{http://www.osnews.com/story/28133/Ford_ditches_Microsoft_for_QNX_in_latest_in-vehicle_tech_platform.html}}.

\item The QNX system of our chosen manufacturer is open-source and enjoys the support of a relatively large third-party developer community.  Currently, no other system is as well established within the automotive sector (although Automotive-Grade-Linux (AGL)\footnote{\url{https://www.automotivelinux.org/announcements/2018/12/17/agl-grows-with-five-new-members}} is increasing in popularity).
\item Our chosen sample-unit allows for bench-testing functionality.  Provided a vehicle can be emulated around the system, it is possible for an investigation to take place in a test environment, whereby only the telematics module (as opposed to the whole vehicle) is required.
\end{itemize}

\subsection{System overview}

Our sample-unit functions identically to more advanced systems from the same manufacturer, but does not support functionality such as voice control or gestures.  (However, code relating to these functions may be found on such units.) The architecture of the sample-unit (illustrated in Figure~\ref{Fig.1})
is divided into two main components: the multimedia service and connected services system, which runs on an X86-based system running QNX; and
an ARM-based system, which manages the CAN-Bus interface that the car uses to communicate with its embedded controllers.

This design made it possible to build a test-bench environment in which the sample-unit's ARM-based module was connected to a vehicle CAN-Bus emulator, thus (to a certain extent) providing a `complete vehicle' environment. We ensured that the emulator was coded with the Vehicle Identification Number (VIN) of the same car from which the sample-unit was taken.

The Intel element of the sample-unit unit is capable of acting as a network gateway with a fixed IP address (see Figure~\ref{Fig.1}). This system allows easy access (through a USB--Ethernet interface) to its internal systems and network. A configuration file allows for several USB interfaces (i.e. an Ethernet to USB converter with firmware matching that of what is presumably used by the manufacturer's technicians) to be used.  Once a suitable USB interface was procured, it was possible to gain the root access password for the Intel system through injecting content into the navigation update service.

With root access enabled, the Intel component allowed execution of processes on both it and the ARM system.  From there, an SSH server could be enabled from which we could login as root using the details procured from the navigation update service.  From here, it became possible to clone the entire file system image of the unit for further analysis.

Using the data recovered, it was possible to build and execute a script containing API information, public and private keys, and login information in order to have the sample-unit send the message content to a local web server set-up on a laptop connected to the vehicle via the USB interface. The data recovered was sanitised and categorised, then used to analyse potential privacy implications.

\begin{figure}[t]
\centering
\includegraphics[scale=0.4]{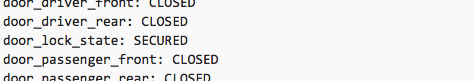}
\caption{An example of captured data containing sensor types and string values.}
\label{Sample}
\end{figure}

\subsection{Data sanitation}

To make better sense of the raw data collected, we went through a data-sanitation process.  The data elements can be described thus.

\begin{itemize}
\item \emph{Obfuscated data.}  This is the raw data as procured through the experiment set-up.  The data as it stands would not be usable for analytical purposes as it has been obfuscated by the manufacturer. In this case, it refers to a message ID of a proprietary value, sensor type and sensor data relating to a particular vehicle. 
\item \emph{Sensor types.}  Of the returned datasets, these would be the first variable that is described.  
See, for example, Figure~\ref{Sample}, where door locks and hinge assemblies are shown to have their own sensors relaying data.
\item \emph{Sensor data.}  This pertains to values associated with sensor type variables.  Some values are straightforward (e.g. ON/OFF, OPEN/CLOSED, and TIME AND DATE), while some require more interpretation.

\item \emph{Informed descriptive data.} To provide relevant meaning to the data collected, a description was added to the variables and values that provided a non-numerical overview of what kind of data the variable pertained to (e.g. the brake positioning sensor indicates it collects data relating to the brake pedal and performance) and to provide meaning to the value that was being collected (e.g. ON/OFF means the ABS sensor is turned on or off).  To ensure that the descriptions were as accurate as possible, we tried to use as many different sources as possible when analysing the data.  

\end{itemize}

\section{Analysis}

\begin{table}[t]
\caption{Privacy policy unique data points} 
\label{tab:long}
\begin{tabular}{llll}
{\bf Information type} & {\bf Unique data points}  & {\bf Information type} & {\bf Unique data points} \\ 
Speed	&	2	&
Steering inputs	&	2	\\
Braking inputs	&	4	&
Accelerator input	&	3	\\
Seat usage	&	2	&
Door/window usage	&	11	\\
Interior climate	&	7	&
Exterior climate	&	3	\\
Engine status	&	6	&
Fuel status	&	3	\\
Lighting controls	&	2	&
Mileage	&	1	\\
Vehicle coordinates	&	1	&
Date and time	&	2	\\
Infotainment usage	&	7	&
Keyfob status	&	2	\\
Vehicle identification & 1 \\
\label{table}
\end{tabular}
\end{table}

The analysis of the sample-unit provided a plethora of information relating to the use of the vehicle, as well as some more personal features surrounding the use of a car.  It also revealed the wider ecosystem in which the vehicle operates. The data was classified into a number of high-level categories (see Table~\ref{table}). The \textit{information types} represent parts of the core driving experience, such as steering, as well as other areas such as infotainment usage, and capturing  time stamps. The \textit{unique data points} represent specific, unique points of data about a category. For example, \textit{door/window usage} has 11 unique data points, with separate messages covering opening status and how far the window is opened. Some data points contain more information than others.  For example, \textit{speed} can be broken down further into brake usage on each individual wheel. The experiment uncovered points of interest within the connected vehicle ecosystem that merit further discussion. The first topic (monetising sensor information) deals directly with the results in Table~\ref{table}; the further topics representing a development of thoughts from both Table~\ref{table} and the data-acquisition process. 

\subsection{Monetising sensor information}

 The sensors from which our chosen sample-unit records data are capable of providing a detailed picture of vehicle usage that could give rise to excessive profiling. For example, individual wheel speed sensors can be used to determine the angle of a corner, and the speed and the forces being applied in that corner. This data can be combined with throttle positioning and brake force application to develop a driver profile.  Also, the telematics can provide data relating to button presses on the in-vehicle system controller, from which it can be determined how often a user combines on-board infotainment use with driving. 

In its privacy policy, our chosen manufacturer states that it shares its data with a newly formed subsidiary that essentially acts as a white-label data-storage service for around 8.5 million cars.  This data is then stored on cloud-based servers and can be used to broker, for example, pay-as-you-drive insurance, whereby the customer pays a fee based on, for example, the amount of miles covered. This is a model already implemented in (for example) the UK under limited mileage policies via specialist providers~\cite{telematicsprototypes}.

Previously, a pay-how-you-drive model was not viable. However, such a model is now eminently possible.  For example, if an individual often carries passengers, and often drives enthusiastically on busy roads where the potential for accidents may be significant, charges may rise.
Previously, these were questions an insurance company might ask to help calculate the risk of a potential customer; now, however, there is the potential to acquire highly detailed driving reports~\cite{joygerla}. 

It has been reported that, by 2025, the market for data types captured from connected cars could be worth almost 33 billion US dollars\footnote{\url{https://www.mckinsey.com/~/media/McKinsey/Industries/Automotive\%20and\%20Assembly/Our\%20Insights/Monetizing\%20car\%20data/Monetizing-car-data.ashx}}. 
It can be assumed that the potential for significant abuse within these models exists. Looking beyond personal privacy interference, statistical inferences based on 
these datasets and form the basis of accident prediction and decision-making that could (in theory) serve to penalise users with specific driving habits.

\subsection{Privacy policies and controls}

Developing privacy policies becomes difficult when they need to be tailored to a wide range of potential specifications: the privacy information for the connected services platform of one manufacturer is more akin to a `If, Then' statement than a standard policy document\footnote{\url{https://myc-profile.bmwgroup.com/api/gateway/contentserver/staticcontent/Angular/gdpr/v2/?target=bmw-browser\#/legal-docs-content?version=2018.05.15&fileName=Bmw_cd_pp_gb-en.json}}.  In addition, policy documentation is not always available from the manufacturer, and in many countries such documentation is not explicitly agreed to upon purchase.  When comparing privacy policies to our data, it becomes clear that users may not be aware of the amount of unique data points that are collected, as none of the policies have been that specific.

Typically, users are not made aware of the ways in which their data may be collected or used, and have no control over who can access their data, how their data is processed, or if it is shared with third parties. In many cases, the user is not made aware of the existence of the privacy policy, especially in cases where an owner purchases the vehicle via the second-hand market. 

\subsection{Third-party applications}

Connected cars are often built upon platforms that allow for the installation and use of third-party applications and services.  Our sample-unit is no different, collecting data on the use of and interaction with these applications (although this functionality was not explicitly tested). In many cases, these applications perform functions analogous to those one can download and use on any mobile device, such as a smartphone. As such, user privacy concerns in these areas mirror those found within mobile applications development and usage. 

Many of these applications are designed to adapt to any given user's specific needs and context.  However, these applications often do not provide mechanisms to provide users control over the kind of data that is collected and used by these applications, thereby giving rise to potential privacy violations~\cite{locationbasedapps}. These applications can also be developed and installed without the manufacturer's knowledge or consent, and therefore are not subject to any controls the manufacturer may have placed on vehicle system access.

It is, therefore, important to highlight the need to provide structural guarantees to users of connected cars in order to provide confidence that data confidentiality is ensured throughout the ecosystem.  Of course, to do this, there is first a need to be aware of users' privacy expectations, as well as what constitutes a trustworthy ecosystem~\cite{protectingcars}. 

\subsection{Data confidentiality within the wider ecosystem}

In~\cite{miorandi} Miorandi \emph{et al.} define data confidentiality to be one of the defining issues faced by those designing and developing IoT systems.  As a consequence of the large volumes of data generated by a connected car over its lifetime, together with the limitations of control over its data transmission systems, current approaches to preserving confidentiality may not be applicable to connected vehicles. 

From our knowledge of the sample-unit (see Figure~\ref{Fig.1}), we can see that connected vehicles are highly reliant on continuous wireless connectivity from third-party service providers --- which are known to be potentially vulnerable to various intrusions, including unauthorised network access, man-in-the-middle attacks, network jamming or interference, spoofing and denial-of-service attacks~\cite{carcomms}.

It is argued in~\cite{suo} that information networks that support IoT applications need to be able to guarantee identification, integrity, confidentiality and undeniability.  From a connected vehicle perspective, it is argued that network availability is the most important factor, followed by confidentiality~\cite{Kasinathan}.  Confidentiality issues arise due to the volume of data generated, as well as the effectiveness of control systems for access to these dynamic data streams~\cite{miorandi}.  There are also issues related to vehicle identity management (discussed further below).  This makes cars as vulnerable to attack as any other IoT device. User privacy and security can become compounded by this lack of data integrity and confidentiality, and unauthorised access to or interference with systems within the car could hamper its ability to function safely \cite{surveyonsecurity}.

\subsection{The automotive lifecycle}

With the lifecycle of an average car being approximately nine years\footnote{\url{http://www.buckingham.ac.uk/wp-content/uploads/2014/11/pnc-2014-usedcar.pdf}}, a connected car has a longer lifespan than the typical IoT device.  In addition, it is significantly more likely to be re-sold over its lifetime. However, from our assessment of the data our chosen sample-unit collects, as well as the manner in which it does so, there does not appear to be an easy means of differentiating between users, so as to potentially generate data that could harm previous users when utilized for for the monetization of sensor data. Therefore, the vehicle continues to collect data as if it were being used by only one person.  Furthermore, due to the fact that the vehicle is primarily identified by its VIN, within our system the possibility existed to continue monitoring the vehicle's use through applications that allow some remote information display or basic remote access. Further potential privacy infractions may occur at the disposal stage, where the vehicle may be recycled, or stripped for parts --- another area where the connected car differs from many other IoT implementations.

\subsection{A lack of standardisation}

An issue that recurred within this study related to accurately defining telematics as a concept within the industry.  A lack of standardisation within components used in automotive telematics systems means it is difficult to ascertain a single definition of telematics within connected cars.  

As there are so many different platforms, components and systems, it becomes difficult to ensure that data confidentiality and long-term availability is maintained for users throughout the supply chain of these products.  From the investigation of our sample-unit, it is by no means a certain prospect that the manufacturer will be able to maintain their infrastructure,
 nor that the systems built into the vehicles will be able to maintain their availability over the lifecycle of the car, which, on average, is significantly longer than the projected lifespan of many other IoT devices. This leads to complexities with regards to designing adequate privacy policies.

\section{Conclusions}

The current state-of-the-art within connected vehicles reveals that a significant amount of work still needs to take place in order to secure these vehicles. The technology within these vehicles has an exponential development rate accompanied by a long usage lifecycle: security, policy and the legality of what is being implemented in many cases needs to catch up with the technological changes.   Furthermore, the academic literature reveals that there is significant scope for improvement in understanding exactly how these vehicles collect and use data.  

We have provided an assessment of privacy-related threats associated with the connected vehicle landscape.  This assessment was supported by an analysis of a popular telematics systems produced by a global manufacturer.  As with any study of this nature, there are limitations to what has been done.  

First, due to the available budget, only a single sample-unit has been procured.  Such systems are designed to not function unless they are installed into a vehicle where all the sub-components have a matching VIN.  In order to overcome this, a bench-testing environment was used, whereby an emulator took on most of the functions that the sample-unit was expected to interface with.  However, this does not generate any simulated vehicle data.  Second, the processes used to generate the messages arise out of a reverse-engineering process, which may have led to some functionality not being captured.  Third, although great care was taken in ensuring that the procured telematics unit represented the largest possible group of connected vehicles, the results may not reflect other manufacturers' approaches

Planned future research activities include performing similar analyses on other types of telematics units, such as those based on AGL, and those from different manufacturers. Also, as this paper serves only as a high-level privacy analysis, there remains significant scope for a more in-depth analysis on the future business models that these datasets enable, as well as attempting to gain a better understanding of the end-users' perceptions of their privacy. 

\bibliographystyle{bibstyle/splncs04}

\bibliography{References}

\end{document}